\documentclass[reprint]{JASAnew}

\usepackage{ulem} % for strikethrough 

\begin{document}

\title[Microsecond Precision from Slow Dynamics]{Microsecond-precision sound localization emerges from slow equilibrium dynamics}
\author{Toshio Irino}
\affiliation{Center for Innovative and Joint Research, Wakayama University, 930 Sakaedani, Wakayama, 640-8510, Japan}
\affiliation{School of Informatics and Engineering, The University of Electro-Communications, 1-5-1 Chofugaoka Street, Chofu, Tokyo, 182-8585, Japan.}
\affiliation{Yamasemi Lab, Nara, Japan}

% \author[a,b,c,1]{Toshio Irino}

\preprint{Toshio Irino, JASA}		%  if you want want this message to appear in upper left corner of title page

\date{\today} 

\begin{abstract}
Precise sound localization relies on microsecond sensitivity to interaural time differences (ITDs), yet binaural perception exhibits sluggish tracking of dynamic acoustic cues. How such extraordinary temporal precision arises despite comparatively slow neural responses remains unresolved. This study proposes that ITD is represented as a stable equilibrium of neural population dynamics rather than by the classical place-coding mechanism based on delay-line coincidence detection. In this framework, excitatory and inhibitory interactions across frequency channels drive the system toward an equilibrium corresponding to the estimated ITD. The resulting dynamics achieve microsecond-level precision and reproduce key physiological observations, including frequency-dependent best-delay distributions, without requiring explicit delay lines or precisely timed inhibition. These results challenge the classical place-coding framework and suggest a fundamentally different principle for binaural computation. More generally, the findings suggest that microsecond-level sensitivity and sluggish binaural perception may be complementary consequences of the same equilibrium dynamics, providing a potential framework for reconciling these seemingly disparate phenomena.
\end{abstract}

%% Keywordsはsubmission pageで入力
%\begin{keywords}
%Binaural hearing； Interaural time difference (ITD);  Population coding;  Place coding;  Neural computation
%\end{keywords}

%% pacs numbers not used

\maketitle

%  End of title page for Preprint option --------------------------------- %

\section{Introduction}
\label{sec:Introduction}

Sound localization in the horizontal plane relies primarily on interaural time differences (ITDs) for low-frequency sounds. Although naturally occurring ITDs typically range from several tens to a few hundred microseconds depending on head size, humans can discriminate changes in sound-source azimuth of approximately 1 degree, corresponding to ITD differences on the order of 10 $\mu\mathrm{s}$. Explaining how such extreme sensitivity arises remains a central challenge for models of binaural hearing.

\begin{figure*}[t]
    \begin{minipage}{0.49\textwidth}
        \centering
 \includegraphics[width=0.8\columnwidth]{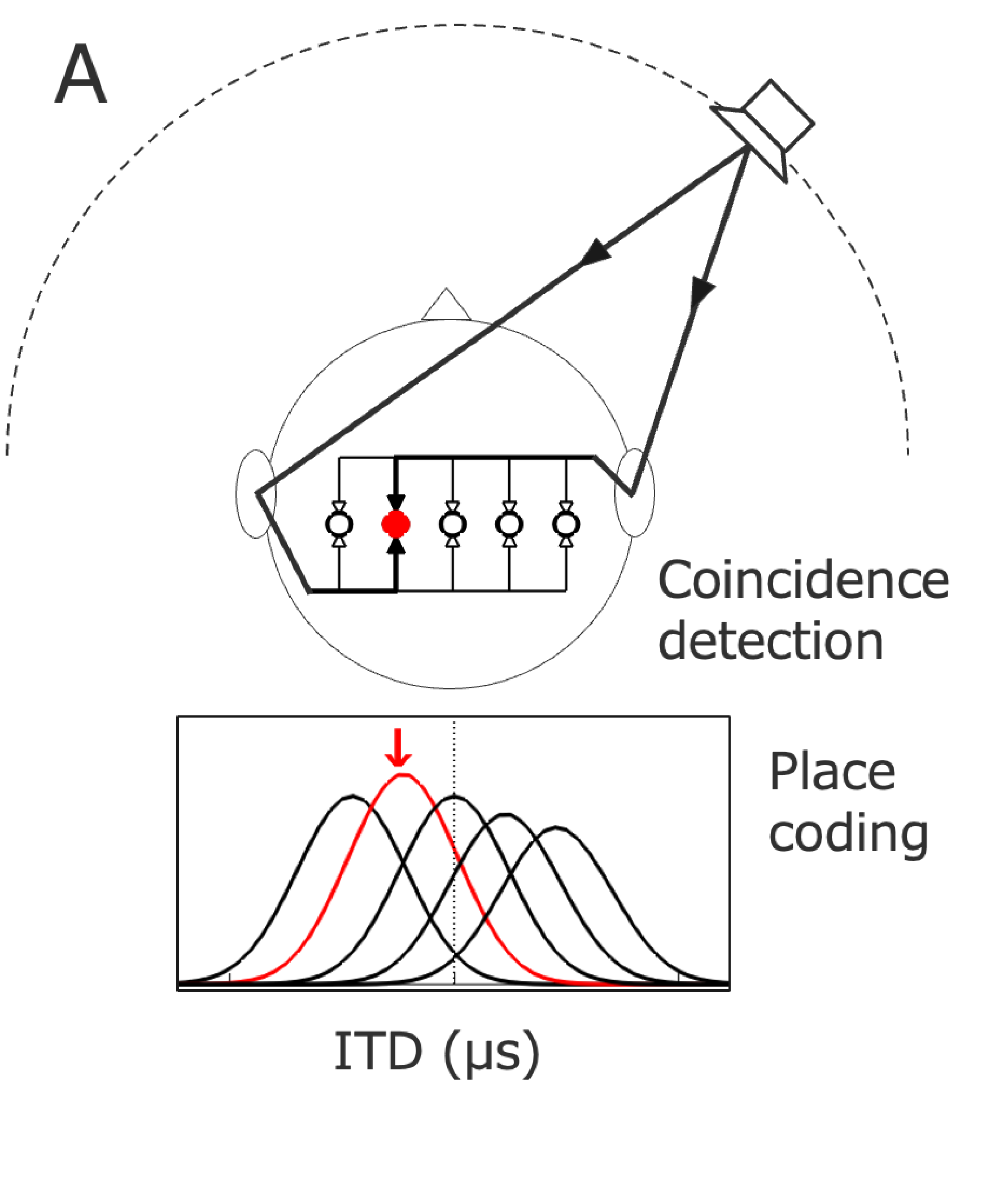}
    \end{minipage}
    \begin{minipage}{0.49\textwidth}
        \centering
      \includegraphics[width=0.85\columnwidth]{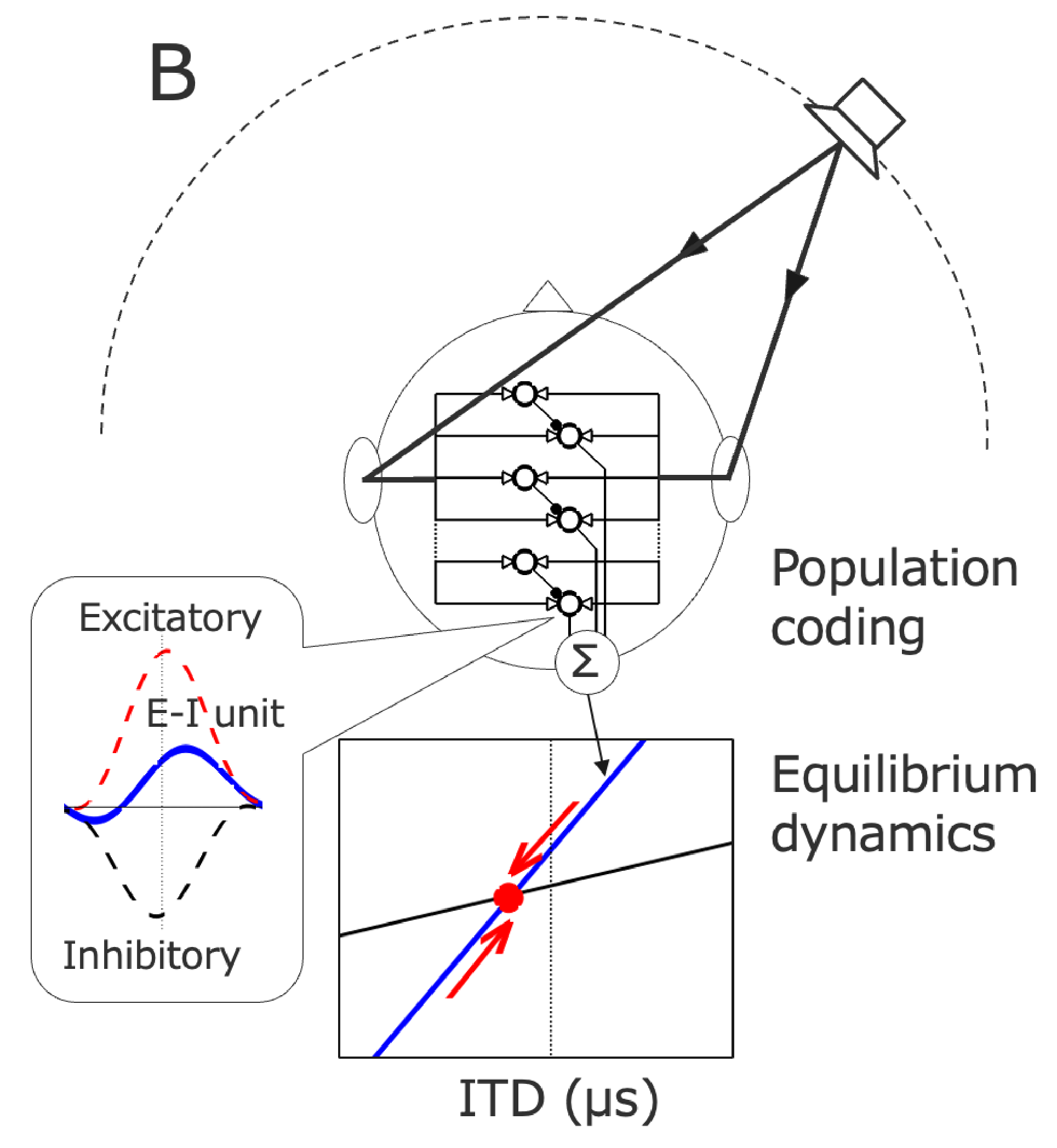}
    \end{minipage}
    
\caption{
Schematic illustration of sound localization based on interaural time differences (ITDs).
\textbf{A.} Jeffress model \cite{jeffress1948place}, based on coincidence detection and place coding.
\textbf{B.} The proposed model, based on population coding and equilibrium dynamics.
A sound source produces a time difference between the two ears, referred to as the ITD.
The figure illustrates how this ITD is processed within neural circuits. The red filled circle represents the estimated ITD of the sound source in each framework.
See the main text for details.
\label{fig:Model_Jeffress_EqlbmDyn}
}

\end{figure*}

For nearly eight decades, explanations of ITD processing have largely been framed within the place-coding architecture proposed by Jeffress in 1948, in which ITD is represented by coincidence detectors supplied by axonal delay lines (Fig.~\ref{fig:Model_Jeffress_EqlbmDyn}A).
Different neurons are maximally activated by different ITD values, thereby forming a spatial map of sound azimuth.
However, while systematic delay lines and spatial maps of ITD have been convincingly demonstrated in avian species~\cite{carr1988axonal,carr1990circuit},
their applicability to mammals has remained under discussion
~\cite{mcalpine2001neural, mcalpine2003sound, grothe2010mechanisms}.
In mammals, anatomical evidence for precisely arranged delay lines is limited, and physiological recordings reveal ITD tuning characteristics that are not readily explained by a straightforward place code
~\cite{joris1998coincidence, joris2007matter}. 
A notable observation is that ITD-sensitive neurons in the mammalian medial superior olive (MSO) often exhibit best delays (BDs) extending beyond the ecologically relevant ITD range imposed by head size~\cite{yin1990interaural,mcalpine2001neural}.
These findings are not readily reconciled with a simple place-based representation of ITD based on single-neuron tuning peaks.
Consistent with this view, physiological studies have demonstrated that precisely timed inhibition plays a critical role in shaping ITD sensitivity in the mammalian MSO~\cite{brand2002precise}. 
Rather than being determined solely by axonal delay lines, ITD tuning has been proposed to arise from the interaction between excitatory and temporally precise inhibitory inputs. However, this interpretation has been challenged, particularly with respect to whether inhibition can provide the temporal precision required for microsecond ITD tuning~\cite{joris2007matter}.

%%----------------------------------
%%% model 
%%----------------------------------
%%----------------------------------

More fundamentally, it remains unresolved how microsecond-level ITD sensitivity arises despite the relatively slow temporal dynamics of binaural perception, a phenomenon commonly referred to as binaural sluggishness \cite{grantham1978detectability,kollmeier1990binaural,hauth2018modeling}. Many psychophysical studies have shown that binaural perception tracks dynamic changes in interaural cues only over time scales of a few tens of milliseconds. Although a more recent study demonstrated that faster modulations can be detected when the stimulus is carefully designed \cite{siveke2008psychophysical}, the temporal resolution remains, at best, on the order of milliseconds, still far from the microsecond precision observed in static ITD sensitivity. In contrast, classical delay-line coincidence-detection models are based on instantaneous neural responses to ITD and provide no obvious explanation for such sluggish perceptual dynamics.

% In contrast, the Jeffress model, based on coincidence detection, predicts instantaneous responses to ITD changes. This apparent paradox motivates the hypothesis that ITD may be represented through a dynamical process operating on longer time scales.

% 50--200~ms

In this study, ITD is proposed to be encoded as a stable equilibrium of a dynamical system driven by neural population activity, as illustrated in Fig.~\ref{fig:Model_Jeffress_EqlbmDyn}B. Within this framework, microsecond-level ITD sensitivity can be achieved despite neural dynamics operating on time scales of tens of milliseconds. Importantly, the proposed model does not rely on the place-coding architecture of Jeffress, nor does it require explicit delay lines or precisely timed inhibitory interactions. 
The following sections show that this framework accounts for key physiological observations and suggest that precise temporal computation may emerge from stable neural dynamics operating on much slower time scales.

%% %%%%%%%%%%%%%%%%%%%%%%%%%%%%%%%%%%%%%%%%%%%%%%%
\section{Equilibrium-Dynamics Model}
%% %%%%%%%%%%%%%%%%%%%%%%%%%%%%%%%%%%%%%%%%%%%%%%%

%% %%%%%%%%%%%%%%%%%%%%%%%%%%%%%%%%%%%%%%%%%%%%%%%
\subsection{Modeling the neural response} \label{sec:ModelNeural}
%% %%%%%%%%%%%%%%%%%%%%%%%%%%%%%%%%%%%%%%%%%%%%%%%
A simplified model was employed to clarify the behavior of the proposed framework. 
The peripheral auditory processing up to the MSO follows conventional extension models of Jeffress-type frameworks \cite{lindemann1986extension,shackleton1992across,breebaart2001binaural}.
For cochlear frequency analysis, the linear passive gammachirp (GC) filterbank \cite{irino1997time,irino2006dynamic} was adopted because the present study focuses exclusively on ITD estimation under conditions in which the sound levels at the two ears are essentially identical. Accordingly, level-dependent cochlear nonlinearities were not included.
The peak frequencies of the filters range from 100~Hz to 1.5~kHz, consistent with bandwidth characteristics reported for guinea pigs~\cite{evans1972frequency}, cats~\cite{liberman1978auditory}, and gerbils~\cite{schmiedt1989spontaneous}.
The sampling rate was set to $f_s = 96$ kHz, yielding a temporal resolution of $10.4~\mu\mathrm{s}$, which is comparable to the ITD difference corresponding to the minimum audible angle ($\sim 1$ degree).

The output of the filterbank is processed by the Meddis hair-cell model \cite{meddis1986simulation}, producing a response representing the probability distribution of auditory nerve firing. 
Although bushy cells in the anteroventral cochlear nucleus (AVCN) are known to enhance phase locking \cite{joris1998coincidence}, this stage is omitted here, as it does not critically affect the behavior examined in this study.

\begin{figure}[t]
 \centering
       \includegraphics[width=1\columnwidth]{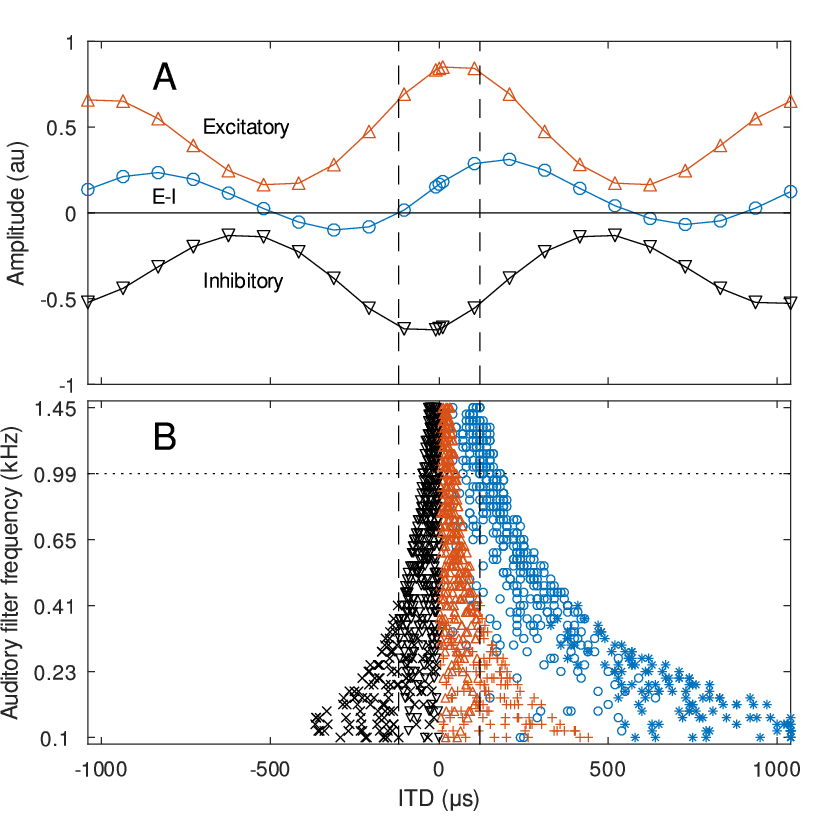}
\caption{ Characteristics of the model units.
    \textbf{A.}  ITD tuning curves of the excitatory unit (red), the inhibitory input (black), and the resulting excitatory--inhibitory (E--I) interaction (blue) for a unit with a best frequency (BF) of 1~kHz.
     \textbf{B.} Best-delay (BD) distributions across auditory-filter frequency for the excitatory unit (red), inhibitory input (black), and the resulting E--I interaction (blue).
    Vertical dashed lines indicate the ecologically relevant ITD range for the gerbil ($\pm120\,\mu\mathrm{s}$).
    Further details are provided in the captions of Fig.~\ref{fig:ITDfunction_BrandFig3c_Model}B and Fig.~\ref{fig:BDdistrRsp}A.
\label{fig:Model+BDdistr}
}
\end{figure}

The responses from the left and right ears ($L_i(t),R_i(t)$) are projected to the MSO, where a running interaural cross-correlation function is computed, as commonly assumed in binaural models \cite{lindemann1986extension,shackleton1992across}:

\begin{equation}
r_{\mathrm{EE}ij}(t; \tau_{\mathrm{BD}ij})
= \int_{-\infty}^{t} L_{i}(s)\cdot  
R_{i}\bigl(s - \tau_{\mathrm{BD}ij}\bigr)\,
e^{-(t-s)/T_{c}} ds .
\label{eq:rEEij}
\end{equation}
where $\tau_{\mathrm{BD}ij}$ denotes the BD of the $j$th unit in the $i$th cochlear channel, and $T_c$ is the integration time constant, set to $200~\mu\mathrm{s}$. 
An example ITD tuning curve of this excitatory unit is shown in red in Fig.~\ref{fig:Model+BDdistr}A.
Up to this point, the formulation is consistent with conventional models
~\cite{lindemann1986extension,shackleton1992across,breebaart2001binaural}.
The key difference lies in the introduction of inhibition. 
It is assumed that inhibitory inputs arrive slightly delayed relative to excitatory inputs and formulate the excitatory--inhibitory (E--I) interaction as:
\begin{equation}
r_{\mathrm{EI}ij}(t) =
r_{\mathrm{EE}ij}(t+\delta; \tau_{\mathrm{BD}ij}) 
- a \cdot r_{\mathrm{EE}ik}(t-\delta; \tau_{\mathrm{BD}ik}) .
\label{eq:rEIij}
\end{equation}
This formulation effectively implements a temporal differencing operation and does not require precisely timed inhibitory inputs. Figure~\ref{fig:Model+BDdistr}A illustrates the ITD tuning curves of the inhibitory input (black) and the resulting E--I interaction (blue) for a unit with a best frequency (BF) of 1~kHz. Hereafter, the output of this interaction is referred to as an E--I unit. The rationale underlying this formulation is described in Section~\ref{sec:SingleNeuron}.

For each cochlear channel with peak frequency $f_{pi}$, the BDs are randomly and independently assigned from uniform distributions over the ranges
\begin{equation}
\bigl\{ \tau_{\mathrm{BD}ij} \mid 0 < \tau_{\mathrm{BD}ij} < \tau_{lim} \cdot \tfrac{1000}{f_{pi}} \bigr\} ~\textrm{and} \nonumber
\end{equation}
\begin{equation}
\bigl\{ \tau_{\mathrm{BD}ik} \mid -\tau_{lim} \cdot \tfrac{1000}{f_{pi}} < \tau_{\mathrm{BD}ik} < 0 \bigr\}
\label{eq:tauBD}
\end{equation}
where $\tau_{lim} = 50~\mu\mathrm{s}$, corresponding to one-twentieth of the period of a sinusoid at the peak frequency $f_{pi}$.
The resulting BD distributions for the excitatory units, inhibitory inputs, and the resulting E--I units are shown in Fig.~\ref{fig:Model+BDdistr}B. This configuration allows the E--I units to exhibit BDs larger than those of the individual excitatory units and introduces a frequency-dependent distribution, consistent with physiological observations~\cite{mcalpine2001neural,mcalpine2005creating,joris2007matter}. Even when both excitatory and inhibitory BDs lie within the ecological range, their interaction can produce effective BDs outside this range. This point will be revisited in Sections~\ref{sec:SingleNeuron} and \ref{sec:FreqDependBD}. 

The assumption in Eq.~\ref{eq:tauBD} is introduced for model construction, and the actual physiological implementation may differ to some extent. However, preserving the relative ordering of BDs, such as $\tau_{\mathrm{BD}ij} > \tau_{\mathrm{BD}ik}$, may be essential because reversing this ordering would invert the derivative-like response of the E--I unit (blue curve in Fig.~\ref{fig:Model+BDdistr}A), causing the resulting E--I units to exhibit large negative BDs.

With this configuration, accurate ITD estimation can be achieved through the following equilibrium dynamics. Importantly, excitatory and inhibitory BDs are assigned randomly, demonstrating that accurate ITD estimation does not require precise or systematically organized delay tuning.

%%----------------------------------
%%% ITD estimation & dynamics 
%%----------------------------------
%%----------------------------------

%% %%%%%%%%%%%%%%%%%%%%%%%%%%%%%%%%%%%%%%%%%%%%%%%
\subsection{ITD estimation through equilibrium dynamics}
%% %%%%%%%%%%%%%%%%%%%%%%%%%%%%%%%%%%%%%%%%%%%%%%%
% The core mechanism of ITD estimation based on the proposed equilibrium dynamics framework is described below. 
The outputs of all E--I units in Eq.~\ref{eq:rEIij} are  combined at the population level:
\begin{equation}
B(t) = \sum_{i,j} r_{\mathrm{EI}ij}(t).
\end{equation}
This activity exhibits both positive and negative fluctuations depending on the input sound, with its amplitude determined by the stimulus type and sound pressure level.
To normalize this activity, a half-wave rectified response is defined and subsequently averaged using an exponential moving average:
\begin{align}
A(t) &= \sum_{i,j} \max\!\bigl(r_{\mathrm{EI}ij}(t),\,0\bigr), \\
\bar{A}(t) &= \frac{1}{T_A} \int_{-\infty}^{t} A(s)\, e^{-(t-s)/T_A} ds,
\label{eq:barA}
\end{align}
where $T_A = 50~\mathrm{ms}$. 
The normalized population response is then given by
\begin{equation}
G(t) = \frac{B(t)}{\bar{A}(t)+\varepsilon},
\label{eq:Gt=}
\end{equation}
where $\varepsilon$ is a small constant preventing division by zero.

The ITD estimate $\hat{\tau}(t)$ is obtained as a dynamical system driven by $G(t)$:
\begin{equation}
\frac{d\hat{\tau}(t)}{dt} =
\eta\,G(t) - \lambda\,\hat{\tau}(t) + c,
\label{eq:dtau/dt=eta}
\end{equation}
where $\eta$ and $\lambda$ represent gain and leak parameters, and $c$ is a bias term.

The discrete-time implementation used for simulation is:
\begin{equation}
\bar{A}(t_{k+1})
= (1-\alpha)\bar{A}(t_k) + \alpha A(t_k),
\quad
\alpha = \frac{\Delta t}{T_A},
\label{eq:A_tk+1}
\end{equation}
\begin{equation}
\hat{\tau}(t_{k+1})
= (1-\lambda)\hat{\tau}(t_k) + \eta G(t_k) + c,
\quad
\lambda = \frac{\Delta t}{T_G},
\label{eq:tau_tk+1}
\end{equation}
with $T_G = 100~\mathrm{ms}$ and $\Delta t = 104~\mu\mathrm{s}$, corresponding to one-tenth of the audio sampling period. In other words, the neural dynamics operate on a much slower time scale than the acoustic signal.

At equilibrium, i.e., when $\frac{d\hat{\tau}(t)}{dt}=0$, the ITD estimate becomes:
\begin{equation}
\hat{\tau} = \frac{\eta}{\lambda} G_{eq} + \frac{c}{\lambda}.
\label{eq:hat_tau=}
\end{equation}
where $G_{eq}$ depends on the input ITD and determines the final ITD estimate.  
This dynamics corresponds to convergence to the equilibrium point defined by the intersection of the population activity curve $y = \eta\,G(t)$ and the ITD line $y = \lambda\,\tau - c$, as indicated by the red filled circle in Fig.~\ref{fig:Model_Jeffress_EqlbmDyn}B.

Importantly, precise ITD estimation does not require microsecond-level precision at individual units, but emerges from population dynamics operating on time scales of tens of milliseconds. Under minimal constraints (Eq.~\ref{eq:tauBD}), appropriate excitatory--inhibitory combinations are integrated across the population to yield a stable ITD estimate. In this framework, ITD is represented as the equilibrium state of a dynamical system rather than as a place-coded variable \cite{jeffress1948place}.

%%----------------------------------
%%% ITD Equilibrium dynamics 
%%----------------------------------
 \begin{figure}[t]
 \centering
     \includegraphics[width=1\columnwidth]{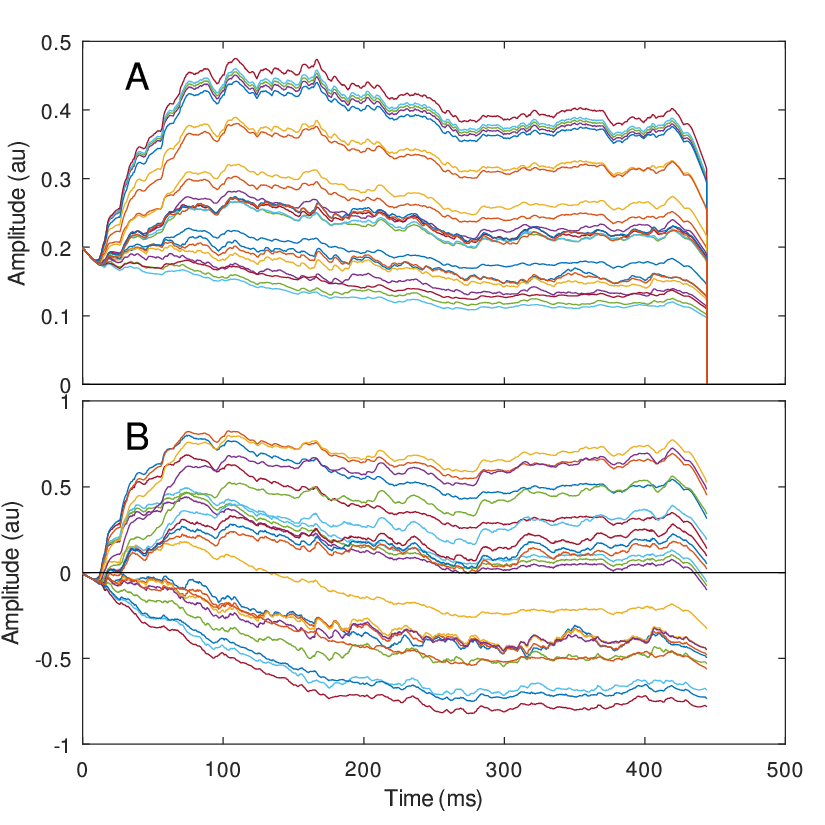}
     \vspace{-20pt}
\caption{ 
Temporal convergence trajectories through equilibrium dynamics for 22 input ITDs ranging from $-1040~\mu\mathrm{s}$ to $1040~\mu\mathrm{s}$.
\textbf{A.} Moving averages of the half-wave-rectified activity $\bar A(t_k)$ defined in Eq.~\ref{eq:A_tk+1}.
\textbf{B.} Internal estimates of $\hat\tau(t_k)$ defined in Eq.~\ref{eq:tau_tk+1} before conversion to ITD values.
\label{fig:AnormTau}
}
\end{figure}  
%%----------------------------------

%%----------------------------------
%%% ITD tuning curve
%%----------------------------------
\begin{figure}[t]
\vspace{-10pt}
 \centering    \includegraphics[width=1\columnwidth]{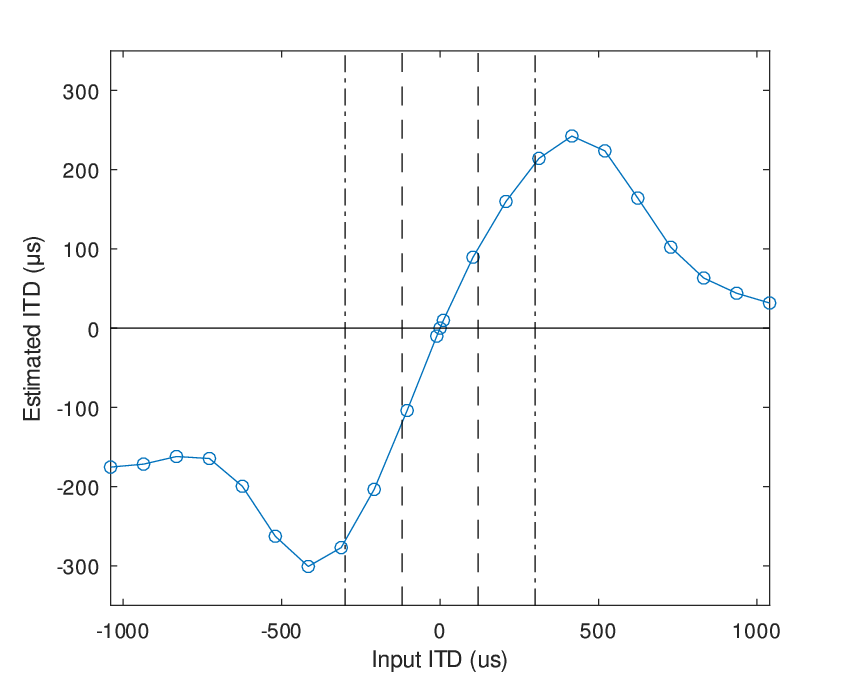}
     \vspace{-20pt}
\caption{
Estimated ITD as a function of input ITD, defined as the mean over the final 50~ms of the trajectories shown in Fig.~\ref{fig:AnormTau}B. The ecologically relevant ITD ranges for the gerbil ($\pm120\,\mu\mathrm{s}$; dashed lines) and guinea pig ($\pm300\,\mu\mathrm{s}$; dash-dotted lines) are also shown.
\label{fig:ITDfunc}
}
\end{figure}  
%%----------------------------------

%% %%%%%%%%%%%%%%%%%%%%%%%%%%%%%%%%%%%%%%%%%%%%%%%
\section{Results}\label{sec:Results}
%% %%%%%%%%%%%%%%%%%%%%%%%%%%%%%%%%%%%%%%%%%%%%%%%
%% %%%%%%%%%%%%%%%%%%%%%%%%%%%%%%%%%%%%%%%%%%%%%%%
\subsection{ITD estimation through equilibrium dynamics}
%% %%%%%%%%%%%%%%%%%%%%%%%%%%%%%%%%%%%%%%%%%%%%%%%

The model was evaluated using bandpass noise (100--1500~Hz) with identical waveforms presented to both ears except for the ITD. The stimulus duration was 440~ms, including 20~ms raised-cosine onset and offset ramps. Input ITDs ranged from $-1040~\mu\mathrm{s}$ to $1040~\mu\mathrm{s}$ in steps of $104~\mu\mathrm{s}$, with additional fine-resolution inputs of $\pm 10.4~\mu\mathrm{s}$. All 22 ITD values were selected as integer multiples of the sampling interval ($10.4~\mu\mathrm{s}$ at a sampling frequency of 96~kHz) to avoid temporal interpolation errors.
Fig.~\ref{fig:AnormTau}A shows the moving averages of the half-wave-rectified activity $\bar A(t_k)$ defined in Eq.~\ref{eq:A_tk+1}. Each line represents the temporal evolution for a different input ITD. Fig.~\ref{fig:AnormTau}B shows the corresponding internal estimates $\hat\tau(t_k)$ defined in Eq.~\ref{eq:tau_tk+1} before conversion to ITD values. The trajectories converge toward distinct equilibrium points corresponding to different input ITDs.

The mean value over the final 50~ms of each trajectory in Fig.~\ref{fig:AnormTau}B was converted to an ITD estimate and plotted against the input ITD in Fig.~\ref{fig:ITDfunc}. The estimated ITD was obtained from the equilibrium value by calibrating the parameters $\eta$ and $c$ in Eqs.~\ref{eq:dtau/dt=eta} and \ref{eq:hat_tau=} so that the estimate matched the input ITD at two reference points (0~$\mu\mathrm{s}$ and $-104~\mu\mathrm{s}$).

The model exhibits approximately linear behavior over the range from $-300~\mu\mathrm{s}$ to $200~\mu\mathrm{s}$, which is substantially wider than the ecologically relevant ITD range of the gerbil ($\pm120~\mu\mathrm{s}$; dashed lines). This approximately linear range is comparable to the ecologically relevant ITD range of the guinea pig ($\pm300~\mu\mathrm{s}$; dash-dotted lines). Beyond this range, the ITD tuning curve becomes progressively shallower and folds back beyond approximately $\pm400~\mu\mathrm{s}$, resulting in a non-monotonic relationship between the estimated and input ITDs.
 
% girbil D = 約 3.2 cm  -->  MaxITD = 120 us

% guinea pig D= 約 4.5 〜 5.5 cm    MaxITD = 約 250 〜 330 µs
% copilot 9 AUg 2026
% Greene et al. (2014) がモルモットの頭部伝達関数 (DTF/HRTF) を測定し、 最大 ITD ≈ 250 μs（低域 DTF から計算） トーンパルスを用いた測定では 320 μs 超

% human D = 約 15 〜 17 cm ~    MaxITD = 約 650 〜 700 µs

Importantly, for input ITDs of $-10.4~\mu\mathrm{s}$ and $10.4~\mu\mathrm{s}$, the estimated ITDs were $-10.0~\mu\mathrm{s}$ and $9.8~\mu\mathrm{s}$, respectively, demonstrating accurate and almost symmetric estimation near zero ITD. These results indicate that precise ITD estimation can be achieved through slow convergence dynamics, as illustrated in Fig.~\ref{fig:AnormTau}B. This property provides a potential resolution to the apparent paradox between microsecond-level ITD sensitivity and limited temporal resolution in dynamic binaural processing.

%Coefficients for mapping Amp to ITDest  =  [359.243,  -19.8623]
%Input ITD [-10.4, 0, 10.4]us --> Estimated ITD [-10.0222,  0,  9.83895] us

%%----------------------------------
%%% Single neuron response
%%----------------------------------
 \begin{figure}[t]
        \centering
        \includegraphics[width=1\columnwidth]{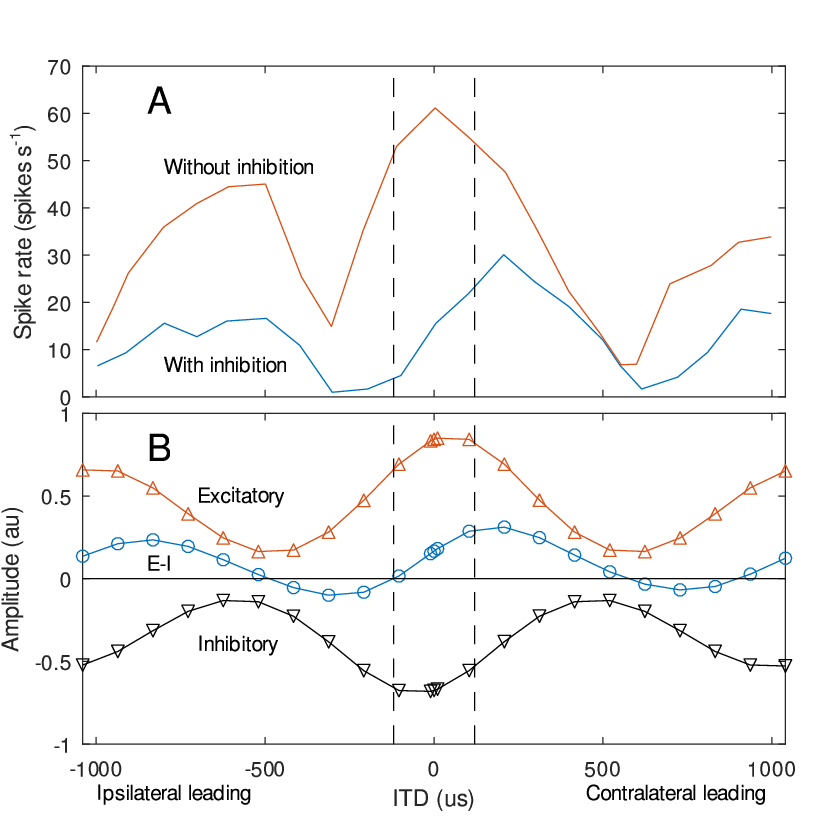}
    \caption{ITD tuning functions.
    \textbf{A.} Physiological ITD tuning of a typical MSO neuron (best frequency \,$\simeq 1\,\mathrm{kHz}$), adapted from Fig.~3c of \citet{brand2002precise}. See the main text for details.
    \textbf{B.} ITD tuning functions of the proposed model units (same as Fig.~\ref{fig:Model+BDdistr}A), shown here for ease of comparison. The BD of the excitatory unit (red) is $+50\,\mu\mathrm{s}$, and that of the inhibitory input (black) is $-50\,\mu\mathrm{s}$. The resulting excitatory--inhibitory (E--I) unit (blue) yields a BD of approximately $180\,\mu\mathrm{s}$. Vertical dashed lines: $\pm 120\,\mu\mathrm{s}$.
\label{fig:ITDfunction_BrandFig3c_Model}}
\end{figure}

%%----------------------------------
%%% BD distribution
%%----------------------------------
 \begin{figure}[t]
        \centering
        \includegraphics[width=1\columnwidth]{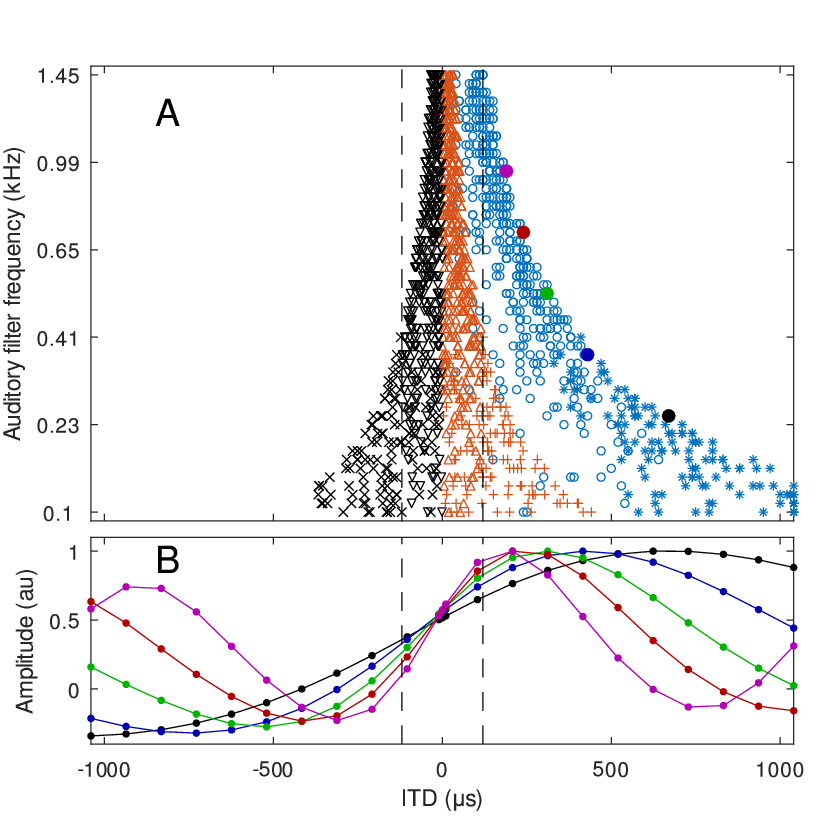}
        \caption{
BD distributions of model units and their ITD tuning curves.
\textbf{A.} Reproduced from Fig.~\ref{fig:Model+BDdistr}B for convenience. Excitatory units (red $\triangle$, $+$), inhibitory units (black $\triangledown$, $\times$), and the resulting E--I units (blue $\circ$, $*$). 
The vertical axis represents the peak frequency of the auditory filter. 
Symbols $\circ$ indicate combinations of $\triangle$ and $\triangledown$ whose BDs both fall within $\pm 120~\mu\mathrm{s}$ (vertical dashed lines), whereas $*$ indicates combinations of $+$ and $\times$ that do not satisfy this condition. Note that BD values exceeding $1040~\mu\mathrm{s}$ are plotted at $1040~\mu\mathrm{s}$.
\textbf{B.} ITD tuning curves of the E--I units corresponding to the colored filled symbols in (A). The peaks are normalized to unity.
\label{fig:BDdistrRsp}
}

\end{figure}   

%% %%%%%%%%%%%%%%%%%%%%%%%%%%%%%%%%%%%%%%%%%%%%%%%
\subsection{Single neuron response}
\label{sec:SingleNeuron}
%% %%%%%%%%%%%%%%%%%%%%%%%%%%%%%%%%%%%%%%%%%%%%%%%

The proposed model is shown to be consistent with physiological observations. Fig.~\ref{fig:ITDfunction_BrandFig3c_Model}A shows the ITD tuning function of a typical MSO neuron with a best frequency of approximately $1\,\mathrm{kHz}$, reported by \citet{brand2002precise}. The blue curve represents the tuning curve measured under normal conditions, exhibiting a BD of approximately $170\,\mu\mathrm{s}$. This value exceeds the ecological ITD range of the gerbil, which is approximately $\pm120\,\mu\mathrm{s}$.

In contrast, when inhibition is pharmacologically blocked by applying the glycine antagonist strychnine, the tuning curve (red line) shifts, resulting in a BD of approximately $50\,\mu\mathrm{s}$, which falls within the ecological range.
These results indicate that inhibition plays a crucial role in shaping ITD tuning and can shift the neuron's preferred delay beyond the range predicted by simple anatomical delays.

The proposed model reproduces this behavior, as shown in Fig.~\ref{fig:ITDfunction_BrandFig3c_Model}B.
The neural response is assumed to arise from the interaction between excitatory and inhibitory inputs (blue curve; $r_{\mathrm{EI}ij}$ in Eq.~\ref{eq:rEIij}).
In the absence of inhibition, the response corresponds to the purely excitatory component (red curve; $r_{\mathrm{EE}ij}$ in Eq.~\ref{eq:rEEij}), whose BD is set to $50\,\mu\mathrm{s}$.
An inhibitory unit with a symmetric BD of $-50\,\mu\mathrm{s}$ (black curve; $r_{\mathrm{EE}ik}$  in Eq.~\ref{eq:rEEij}) is further assumed.
The interaction between the excitatory and inhibitory components effectively shifts the peak of the response function, yielding an E--I response with a BD of approximately $180\,\mu\mathrm{s}$, consistent with the physiological observations shown in Fig.~\ref{fig:ITDfunction_BrandFig3c_Model}A. This indicates that large apparent BDs can emerge even when the individual component BDs lie within the ecological range.
Because inhibitory inputs of this type are difficult to isolate in physiological experiments, the role of such mechanisms may have remained largely overlooked in previous studies.

%\textcolor{red}{--- workmark 9 Aug 2026 ===========}

%% %%%%%%%%%%%%%%%%%%%%%%%%%%%%%%%%%%%%%%%%%%%%%%%
\subsection{Frequency-dependent BD distributions}
\label{sec:FreqDependBD}
%% %%%%%%%%%%%%%%%%%%%%%%%%%%%%%%%%%%%%%%%%%%%%%%%

Fig.~\ref{fig:BDdistrRsp}A reproduces the BD distributions shown in Fig.~\ref{fig:Model+BDdistr}B, where ten E--I units are assumed for each auditory-filter channel. 
Through combinations of excitatory and inhibitory units, the resulting E--I units exhibit BDs that extend beyond the ecological range of $\pm 120\,\mu\mathrm{s}$. Notably, even when both the excitatory and inhibitory unit BDs fall within the ecological range ($\triangle$, $\triangledown$), the resulting E--I unit BD ($\circ$) can lie well outside this range. When this condition is not satisfied ($+$ and $\times$), the resulting E--I unit BD ($*$) can become even larger.
Fig.~\ref{fig:BDdistrRsp}B shows the ITD tuning curves of E--I units corresponding to the colored filled circles in Fig.~\ref{fig:BDdistrRsp}A. These results are consistent with physiological observations reported in previous studies \cite{mcalpine2001neural, mcalpine2005creating, joris2007matter}. 

The single-unit and population-level results together indicate that the distribution of BDs in E--I units can naturally span a wide range of ITDs and exhibit frequency dependence. Importantly, the mapping between excitatory and inhibitory unit BDs does not require precise pairing; as long as a simple sign constraint is satisfied, random combinations are sufficient. Consequently, accurate ITD estimation can emerge without fine tuning at the level of individual units. This provides a simple and robust explanation for the broad distribution of BDs observed in physiological studies, without invoking the precisely arranged delay lines assumed by place-coding models.

%%----------------------------------
%%% Corr BP noise
%%----------------------------------
 \begin{figure}[t]
        \centering
        \includegraphics[width=1\columnwidth]{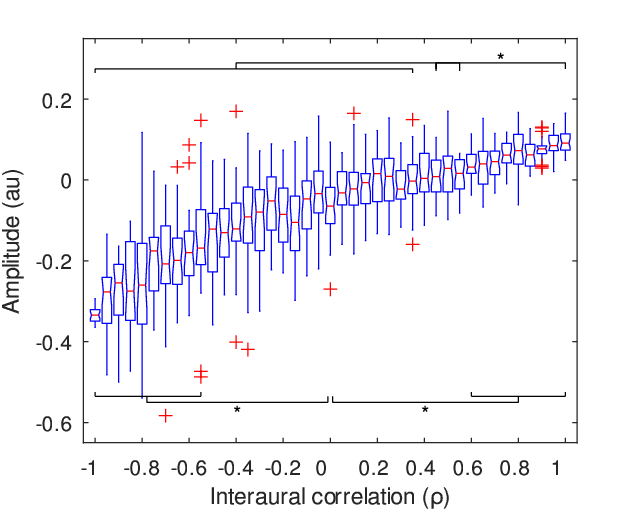}
    \caption{ Boxplots of equilibrium values as a function of interaural correlation ($\rho$). A one-way ANOVA revealed a significant effect of $\rho$. Asterisks indicate significant pairwise differences identified by Tukey's HSD post hoc test ($\alpha = 0.05$). Bartlett's test rejected the assumption of homogeneity of variance. See text for details. \label{fig:BPNoiseCorr}
} 
\end{figure}

%% %%%%%%%%%%%%%%%%%%%%%%%%%%%%%%%%%%%%%%%%%%%%%%%
\subsection{Sensitivity to interaural correlation}
\label{sec:SenseInterCorr}
%% %%%%%%%%%%%%%%%%%%%%%%%%%%%%%%%%%%%%%%%%%%%%%%%

~\citet{joris2007matter} argued that a plausible model of ITD processing should account not only for ITD estimation itself but also for binaural sensitivity to interaural correlation ($\rho$). 
In fact,  \citet{gabriel1981interaural} reported remarkably high human sensitivity to changes in interaural correlation. For narrowband noise, the just noticeable difference (JND) in interaural correlation ($\Delta\rho$) was found to be smaller than 0.01 and to increase with stimulus bandwidth. Any biologically plausible model of binaural processing should therefore be capable of supporting such sensitivity.

To examine this issue, the response of the proposed model was evaluated as a function of $\rho$. The value of $\rho$ was varied from $+1$ to $-1$ in steps of 0.05. The stimulus consisted of bandpass noise spanning 100--1000 Hz, generated by summing sinusoidal components at 1-Hz intervals with random initial phases. The equilibrium value was calculated as the mean of the final 50~ms of the internal estimate $\hat\tau(t_k)$ in Eq.~\ref{eq:tau_tk+1}, before conversion to calibrated ITD values. For each value of $\rho$, 30 equilibrium values were obtained from independently generated noise realizations.

The results are shown in Fig.~\ref{fig:BPNoiseCorr}. The variance of the estimated ITD was small for $\rho=\pm1$ and substantially larger near $\rho=0$. In general, the variance increased as the absolute value of $\rho$ decreased. This behavior is consistent with the reduced ability to localize sound sources when the binaural signals become less correlated. 
The observed dependence of response variability on $\rho$ may therefore be related to the perceptual distinction between a sharply localized sound source and a diffuse spatial percept.

A one-way ANOVA revealed a significant effect of $\rho$ ($F_{40,1189}=58.47$, $p<0.001$).
Tukey's HSD multiple-comparison test ($\alpha=0.05$) was subsequently performed. In Fig.~\ref{fig:BPNoiseCorr}, only comparisons that differed significantly from either $\rho=1$ or $\rho=0$ are shown. Significant differences were observed across multiple pairs of $\rho$ conditions, suggesting a systematic dependence of the model output on interaural correlation.
Bartlett's test rejected the assumption of homogeneity of variance ($\chi^2(40)=470.11$, $p<0.001$).
Taken together, these results show that both the mean and variance of the model response depend on interaural correlation.
These findings are consistent with the challenge raised by \citet{joris2007matter} that binaural processing mechanisms should account for interaural correlation in addition to ITD estimation.

Gabriel and Colburn \cite{gabriel1981interaural} reported that, for band-limited noise with an upper frequency near 1~kHz, the JND was approximately $\Delta\rho \simeq 0.02$ when the reference sound had $\rho=1$, whereas values below $\Delta\rho \simeq 0.3$ were observed when referenced to $\rho=0$. The results in Fig.~\ref{fig:BPNoiseCorr} are not consistent with these psychophysical observations.
The pair of $\rho$ values with the smallest significant difference identified by Tukey's HSD test does not correspond directly to the reported perceptual thresholds. In addition, JNDs corresponding to 75\% correct performance, estimated from Gaussian approximations of the equilibrium-value distributions, did not fully account for the psychophysical data.
This discrepancy suggests that perceptual sensitivity is not determined solely by the overlap of the equilibrium-value distributions. Future work will therefore require an explicit observer model linking the neural representations generated by the present dynamical model to perceptual judgments and testing whether the model can quantitatively account for a broader range of binaural psychophysical findings.

%% %%%%%%%%%%%%%%%%%%%%%%%%%%%%%%%%%%%%%%%%%%%%%%% 
\section{Discussion}\label{sec:Discussion}
%% %%%%%%%%%%%%%%%%%%%%%%%%%%%%%%%%%%%%%%%%%%%%%%%
The classical Jeffress model explains ITD coding solely through excitatory coincidence detection and does not require inhibition. However, the presence of inhibitory inputs in the MSO has been well established \cite{yin1990interaural}, leading to extensions of the Jeffress framework that incorporate inhibition \cite{lindemann1986extension,breebaart2001binaural}. Despite these modifications, these models retain the fundamental place-coding principle, in which ITD is represented by the spatial pattern of activity across neurons tuned to different delays.

In contrast, the proposed model relies essentially on inhibition and represents ITD through population coding governed by equilibrium dynamics. Within this framework, binaural sluggishness arises naturally from the slow convergence of the underlying dynamics. Thus, unlike place-coding models in which localization is determined directly from instantaneous activity patterns, the proposed model provides a mechanism through which slow neural dynamics can coexist with microsecond-level ITD sensitivity.

\citet{joris2007matter} have raised several important issues regarding existing mechanisms of ITD estimation. 
First, they questioned the hypothesis proposed by ~\citet{brand2002precise} that precisely timed inhibition is required for ITD coding. In contrast, the present model does not require dynamically generated precise timing. Instead, it assumes only a slight temporal delay of inhibitory inputs relative to excitatory inputs, together with fixed but randomly distributed BDs. Consequently, microsecond-level ITD sensitivity can emerge from slow population dynamics without requiring microsecond-level precision at the level of individual units.

Second, the Jeffress model assumes that BDs lie within the ecological range and are largely independent of best frequency. However, physiological observations in mammals reveal substantial frequency dependence in BD distributions, with broader distributions at lower frequencies \cite{mcalpine1996interaural,mcalpine2001neural,mcalpine2005creating,joris2007matter}. The proposed model naturally reproduces this characteristic through excitatory--inhibitory interactions (Fig.~\ref{fig:BDdistrRsp}), suggesting that the broad physiological distribution of BDs may emerge from excitatory--inhibitory interactions rather than requiring individual neurons to possess correspondingly large intrinsic delays.

Third, an influential alternative to place coding is the two-channel model \cite{grothe2010mechanisms}, in which ITD is represented by the relative activity of two broadly and oppositely tuned neural populations in the two brain hemispheres. \citet{joris2007matter} pointed out that a drawback of this framework is its reliance on integration across hemispheres, for which lesion studies offer little support. In contrast, the present framework allows accurate ITD estimation within a unilateral neural circuit, suggesting that explicit hemispheric comparison may not be required for the core computation of ITD.

Finally, \citet{joris2007matter} emphasized that binaural processing is important not only for ITD estimation but also for representing interaural correlation in complex acoustic environments. They further noted that neurons with tuning functions resembling the blue curve in Fig.~\ref{fig:ITDfunction_BrandFig3c_Model}A would be expected to show little sensitivity to changes in interaural correlation ($\rho$) near ITD = 0. As described in Section~\ref{sec:SenseInterCorr}, the proposed model exhibited systematic changes in both the mean and variance of its response as a function of $\rho$. Taken together, these findings suggest that the proposed model provides a unified framework that is consistent with the major issues raised by \citet{joris2007matter}.

% \subRelated work  -- eye position / head direction}

The present model computes ITD through dynamic convergence to an equilibrium point.
Similar computational principles have been identified in other neural systems, such as eye-position control \cite{seung1996continuous,dayan2001theoretical} and head-direction representation \cite{zhang1996representation}.
In these systems, variables are represented not by place coding but as stable states of neural dynamics.
Importantly, these functions and ITD-based sound localization share a common computational objective: estimating behaviorally relevant spatial variables that may be critical for survival, such as the locations of prey or predators.
The emergence of similar computational strategies across these diverse neural systems suggests that equilibrium dynamics may provide a general neural mechanism for representing spatial variables.

%% %%%%%%%%%%%%%%%%%%%%%%%%%%%%%%%%%%%%%%%%%%%%%%%
\section{Conclusion}
\label{sec:conclusion}
%% %%%%%%%%%%%%%%%%%%%%%%%%%%%%%%%%%%%%%%%%%%%%%%%
This study proposed an equilibrium-dynamics framework for ITD estimation in which ITD is represented as the equilibrium state of neural population dynamics.
Unlike classical place-coding models based on delay-line coincidence detection \cite{jeffress1948place}, the proposed framework computes ITD through the dynamic convergence of excitatory--inhibitory population activity. Despite relying on relatively slow dynamics, the model achieved microsecond-level ITD estimation and reproduced key physiological observations, including broad and frequency-dependent distributions of BDs. 
More generally, the results suggest that highly precise temporal information can emerge from stable neural dynamics operating on much slower time scales.

%% %%%%%%%%%%%%%%%%%%%%%%%%%%%%%%%%%%%%%%%%%%%%%%%
\begin{acknowledgments}
This research was supported by JSPS KAKENHI: Grant Number JP24K02961. 
The author would like to thank Shiro Suzuki and Rina Kotani for valuable discussions.
\end{acknowledgments}
%% %%%%%%%%%%%%%%%%%%%%%%%%%%%%%%%%%%%%%%%%%%%%%%%

%% %%%%%%%%%%%%%%%%%%%%%%%%%%%%%%%%%%%%%%%%%%%%%%%

\section*{Declarations} 

\subsection*{Conflict of Interest} 
The author declares no conflicts of interest.
\vspace{-1em}

\subsection*{Ethics Approval} 
Not applicable. 
\vspace{-1em}

\subsection*{Data Availability} 
The source code used in this study will be made publicly available through the author's GitHub repository (\url{https://github.com/amlab-wakayama/}) upon publication. No experimental datasets were generated in this study.
\vspace{-1em}

\subsection*{AI Use Statement}
AI-assisted language tools were used during manuscript preparation to improve readability and clarity. All generated content was reviewed, edited, and verified by the author. No AI system contributed to the scientific conception, mathematical derivations, analysis, or conclusions of this work.

%% %%%%%%%%%%%%%%%%%%%%%%%%%%%%%%%%%%%%%%%%%%%%%%%
 %   Appendix  (optional)
%% %%%%%%%%%%%%%%%%%%%%%%%%%%%%%%%%%%%%%%%%%%%%%%%
%\appendix
%\section{Appendices}

%% %%%%%%%%%%%%%%%%%%%%%%%%%%%%%%%%%%%%%%%%%%%%%%%
% ---- Rerefence -------
%% %%%%%%%%%%%%%%%%%%%%%%%%%%%%%%%%%%%%%%%%%%%%%%%
\section*{References}
\vspace{-3em}
\bibliography{Reference_17Jul26}

@article{irino1997time,
    title = {{A time domain, level-dependent auditory filter: the gammachirp}},
    year = {1997},
  journal={J. Acoust. Soc. Am.},
    author = {Irino, Toshio and Patterson, Roy D.},
    number = {1},
    pages = {412--419},
    volume = {101},
    isbn = {9784339013238},
    doi = {10.1121/1.417975},
    issn = {00014966}
}

@article{meddis1986simulation,
  title={Simulation of mechanical to neural transduction in the auditory receptor},
  author={Meddis, Ray},
  journal={J. Acoust. Soc. Am.},
  volume={79},
  number={3},
  pages={702--711},
  year={1986},
  publisher={Acoustical Society of America},
  doi={10.1121/1.393460}
}

@article{irino2006dynamic,
  title = {{A dynamic compressive gammachirp auditory filterbank.}},
  author={Irino, Toshio and Patterson, Roy D},
  journal={IEEE Trans. Audio Speech Lang. Process.},
  volume={14},
  number={6},
    pages = {2222--2232},
    year={2006},
    doi = {10.1109/TASL.2006.874669},
    issn = {1558-7916},
    pmid = {19330044},
  publisher={IEEE}
}

@article{jeffress1948place,
  author  = {Jeffress, Lloyd A.},
  title   = {A place theory of sound localization},
  journal = {Journal of Comparative and Physiological Psychology},
  volume  = {41},
  number  = {1},
  pages   = {35--39},
  year    = {1948},
  doi = {10.1037/h0061495}
}

@article{carr1988axonal,
  author  = {Carr, Catherine E. and Konishi, Masakazu},
  title   = {Axonal delay lines for time measurement in the owl's brainstem},
  journal = {Proceedings of the National Academy of Sciences},
  volume  = {85},
  number  = {21},
  pages   = {8311--8315},
  year    = {1988},
  doi = {10.1073/pnas.85.21.8311}
}

@article{carr1990circuit,
  author  = {Carr, Catherine E. and Konishi, Masakazu},
  title   = {A circuit for detection of interaural time differences in the brainstem of the barn owl},
  journal = {Journal of Neuroscience},
  volume  = {10},
  number  = {10},
  pages   = {3227--3246},
  year    = {1990},
  doi = {10.1523/JNEUROSCI.10-10-03227.1990}
}

@article{yin1990interaural,
  author  = {Yin, Tom C. and Chan, Joseph C. },
  title   = {Interaural time sensitivity in medial superior olive of cat},
  journal = {Journal of Neurophysiology},
  volume  = {64},
  number  = {2},
  pages   = {465--488},
  year    = {1990},
  doi = {10.1152/jn.1990.64.2.465},
}

@article{joris1998coincidence,
  title={{Coincidence detection in the auditory system: 50 years after Jeffress}},
  author={Joris, Philip X and Smith, Philip H and Yin, Tom CT},
  journal={Neuron},
  volume={21},
  number={6},
  pages={1235--1238},
  year={1998},
  publisher={Elsevier},
  DOI = {10.1016/S0896-6273(00)80643-1}
}

@article{joris2007matter,
  author  = {Joris, Philip X. and Yin, Tom C. T.},
  title   = {A matter of time: internal delays in binaural processing},
  journal = {Trends in Neurosciences},
  volume  = {30},
  number  = {2},
  pages   = {70--78},
  year    = {2007},
  DOI = {10.1016/j.tins.2006.12.004}
}

@article{mcalpine2001neural,
  author  = {McAlpine, David and Jiang, Daliang and Palmer, Alan R.},
  title   = {A neural code for low-frequency sound localization in mammals},
  journal = {Nature Neuroscience},
  volume  = {4},
  number  = {4},
  pages   = {396--401},
  year    = {2001},
  doi = {10.1038/86049}
}

@article{brand2002precise,
  title={Precise inhibition is essential for microsecond interaural time difference coding},
  author={Brand, Antje and Behrend, Oliver and Marquardt, Torsten and McAlpine, David and Grothe, Benedikt},
  journal={Nature},
  volume={417},
  number={6888},
  pages={543--547},
  year={2002},
  publisher={Nature Publishing Group UK London},
  doi = {10.1038/417543a}
}

@article{mcalpine2003sound,
  author  = {McAlpine, David and Grothe, Benedikt},
  title   = {Sound localization and delay lines -- do mammals fit the model?},
  journal = {Trends in Neurosciences},
  volume  = {26},
  number  = {7},
  pages   = {347--350},
  year    = {2003},
  doi = {10.1016/S0166-2236(03)00140-1}
}

@article{grothe2010mechanisms,
  author  = {Grothe, Benedikt and Pecka, Michael and McAlpine, David},
  title   = {Mechanisms of sound localization in mammals},
  journal = {Physiological Reviews},
  volume  = {90},
  number  = {3},
  pages   = {983--1012},
  year    = {2010},
  doi={10.1152/physrev.00026.2009}
}

@article{mcalpine2005creating,
  title={Creating a sense of auditory space},
  author={McAlpine, David},
  journal={The Journal of physiology},
  volume={566},
  number={1},
  pages={21--28},
  year={2005},
  publisher={Wiley Online Library},
  doi = {10.1113/jphysiol.2005.083113 }
}

@article{zhang1996representation,
  author  = {Zhang, Kechen},
  title   = {Representation of spatial orientation by the intrinsic dynamics of the head-direction cell ensemble: A Theory},
  journal = {Journal of Neuroscience},
  volume  = {16},
  number  = {6},
  pages   = {2112--2126},
  year    = {1996},
  doi = {10.1523/JNEUROSCI.16-06-02112.1996}
}

@article{lindemann1986extension,
  title={{Extension of a binaural cross-correlation model by contralateral inhibition. I. Simulation of lateralization for stationary signals}},
  author={Lindemann, Werner},
  journal={The Journal of the Acoustical Society of America},
  volume={80},
  number={6},
  pages={1608--1622},
  year={1986},
  publisher={Acoustical Society of America},
  doi = {10.1121/1.394325}
}

@article{breebaart2001binaural,
  title={{Binaural processing model based on contralateral inhibition. I. Model structure}},
  author={Breebaart, Jeroen and Van De Par, Steven and Kohlrausch, Armin},
  journal={The Journal of the Acoustical Society of America},
  volume={110},
  number={2},
  pages={1074--1088},
  year={2001},
  publisher={Acoustical Society of America},
    doi = {10.1121/1.1383297}
}

@article{shackleton1992across,
  title={Across frequency integration in a model of lateralization},
  author={Shackleton, Trevor M and Meddis, Ray and Hewitt, Michael J},
  journal={The Journal of the Acoustical Society of America},
  volume={91},
  number={4},
  pages={2276--2279},
  year={1992},
  publisher={Acoustical Society of America},
    doi = {10.1121/1.403663}
}

@article{grantham1978detectability,
  title={Detectability of varying interaural temporal differences},
  author={Grantham, D Wesley and Wightman, Frederic L},
  journal={The Journal of the Acoustical Society of America},
  volume={63},
  number={2},
  pages={511--523},
  year={1978},
  publisher={Acoustical Society of America},
  doi = {10.1121/1.381751}
}

@article{kollmeier1990binaural,
  author = {Kollmeier, Birger and Gilkey, Robert H.},
  title = {Binaural forward and backward masking: Evidence for sluggishness in binaural detection},
  journal = {Journal of the Acoustical Society of America},
  volume = {87},
  number = {4},
  pages = {1709--1719},
  year = {1990},
  doi = {10.1121/1.399419}
}

@article{hauth2018modeling,
  author = {Hauth, Christopher F. and Brand, Thomas},
  title = {Modeling sluggishness in binaural unmasking of speech for maskers with time-varying interaural phase differences},
  journal = {Trends in Hearing},
  volume = {22},
  pages = {2331216517753547},
  year = {2018},
  doi = {10.1177/2331216517753547}
}

@article{siveke2008psychophysical,
author = {Siveke, Ida and Ewert, Stephan D. and Grothe, Benedikt and Wiegrebe, Lutz},
title = {Psychophysical and Physiological Evidence for Fast Binaural Processing},
journal = {Journal of Neuroscience},
volume = {28},
number = {9},
pages = {2043--2052},
year = {2008},
doi = {10.1523/JNEUROSCI.4488-07.2008}
}

@book{dayan2001theoretical,
  author    = {Dayan, Peter and Abbott, L. F.},
  title     = {Theoretical Neuroscience: Computational and Mathematical Modeling of Neural Systems},
  publisher = {MIT Press},
  year      = {2001},
  address = {Cambridge, Massachusetts; London, England},
  ISBN = {978-0262541855}
}

@article{seung1996continuous,
  author  = {Seung, H. Sebastian},
  title   = {How the brain keeps the eyes still},
  journal = {Proceedings of the National Academy of Sciences},
  volume  = {93},
  number  = {23},
  pages   = {13339--13344},
  year    = {1996},
  doi     = {10.1073/pnas.93.23.13339}
}

@article{evans1972frequency,
  author  = {Evans, Edwin F.},
  title   = {The frequency response and other properties of single fibers in the guinea-pig cochlear nerve},
  journal = {Journal of Physiology},
  volume  = {226},
  pages   = {263--287},
  year    = {1972},
  doi = {10.1113/jphysiol.1972.sp009984}
}

@article{liberman1978auditory,
  title={Auditory-nerve response from cats raised in a low-noise chamber},
  author={Liberman, M Charles},
  journal={The Journal of the Acoustical Society of America},
  volume={63},
  number={2},
  pages={442--455},
  year={1978},
  publisher={Acoustical Society of America},
  doi = {10.1121/1.381736}
}

@article{schmiedt1989spontaneous,
  title={Spontaneous rates, thresholds and tuning of auditory-nerve fibers in the gerbil: comparisons to cat data},
  author={Schmiedt, Richard A},
  journal={Hearing research},
  volume={42},
  number={1},
  pages={23--35},
  year={1989},
  publisher={Elsevier},
  doi = {10.1016/0378-5955(89)90115-9}
}

@article{gabriel1981interaural,
  title={Interaural correlation discrimination: I. Bandwidth and level dependence},
  author={Gabriel, Kaigham J and Colburn, H Steven},
  journal={The Journal of the Acoustical Society of America},
  volume={69},
  number={5},
  pages={1394--1401},
  year={1981},
  publisher={Acoustical Society of America},
  doi={10.1121/1.385821}
}

@article{mcalpine1996interaural,
  title={Interaural delay sensitivity and the classification of low best-frequency binaural responses in the inferior colliculus of the guinea pig},
  author={McAlpine, David and Jiang, Dan and Palmer, Alan R},
  journal={Hearing research},
  volume={97},
  number={1-2},
  pages={136--152},
  year={1996},
  publisher={Elsevier},
  doi={10.1016/S0378-5955(96)80015-3}
}

\end{document}